\documentclass[a4,12pt]{article}
\usepackage{latexsym}
\usepackage{comment}
\usepackage{mathrsfs}
\usepackage[dvipdfmx]{graphicx}
\usepackage{amsmath,amssymb} 
\let\du=\du                     
        
\def\a{\alpha}
\def\b{\beta}
\def\c{\chi}
\def\d{\delta}

\def\f{\phi}
\def\g{\gamma}

\def\k{\kappa}
\def\l{\lambda}
\def\m{\mu}
\def\n{\nu}

\def\p{\pi}

\def\D{\Delta}

\def\L{\Lambda}

\def\cv{{\cal V}}

\def\bo{{\raise-.3ex\hbox{\large$\Box$}}}               
\def\pa{\partial}                                       
\def\TH{{\raise.2ex\hbox{$\displaystyle \bigodot$}\mskip-4.7mu \llap H \;}}
\def\face{{\raise.2ex\hbox{$\displaystyle \bigodot$}\mskip-2.2mu \llap {$\ddot
        \smile$}}}                                      

\def\Tilde#1{\widetilde{#1}}                    
\def\VEV#1{\left\langle #1\right\rangle}        
\def\leftrightarrowfill{$\mathsurround=0pt \mathord\leftarrow \mkern-6mu
        \cleaders\hbox{$\mkern-2mu \mathord- \mkern-2mu$}\hfill
        \mkern-6mu \mathord\rightarrow$}
\def\dvec#1{\vbox{\ialign{##\crcr
        \leftrightarrowfill\crcr\noalign{\kern-1pt\nointerlineskip}
        $\hfil\displaystyle{#1}\hfil$\crcr}}}           

\def\frac#1#2{{\textstyle{#1\over\vphantom2\smash{\raise.20ex
        \hbox{$\scriptstyle{#2}$}}}}}                   
\def\sfrac#1#2{{\vphantom1\smash{\lower.5ex\hbox{\small$#1$}}\over
        \vphantom1\smash{\raise.4ex\hbox{\small$#2$}}}} 
\def\bfrac#1#2{{\vphantom1\smash{\lower.5ex\hbox{$#1$}}\over
        \vphantom1\smash{\raise.3ex\hbox{$#2$}}}}       
\def\afrac#1#2{{\vphantom1\smash{\lower.5ex\hbox{$#1$}}\over#2}}    

\def\[{\lfloor{\hskip 0.35pt}\!\!\!\lceil}
\def\]{\rfloor{\hskip 0.35pt}\!\!\!\rceil}

\def\du#1#2{_{#1}{}^{#2}}

\def\fracmm#1#2{{{#1}\over{#2}}}

\def\low#1{{\raise -3pt\hbox{${\hskip 0.75pt}\!_{#1}$}}}

\def\Tilde#1{{\widetilde{#1}}\hskip 0.015in}

\newskip\humongous \humongous=0pt plus 1000pt minus 1000pt
\def\caja{\mathsurround=0pt}
\def\eqalign#1{\,\vcenter{\openup2\jot \caja
        \ialign{\strut \hfil$\displaystyle{##}$&$
        \displaystyle{{}##}$\hfil\crcr#1\crcr}}\,}
\newif\ifdtup

\newcommand{\be}{\begin{equation}}
\newcommand{\ee}{\end{equation}}
\newcommand{\nbe}{\begin{equation*}}
\newcommand{\nee}{\end{equation*}}

\newcommand{\lb}{\label}

\def\Tilde#1{\widetilde{#1}}                    

\def\caja{\mathsurround=0pt}
\def\eqalign#1{\,\vcenter{\openup2\jot \caja
        \ialign{\strut \hfil$\displaystyle{##}$&$
        \displaystyle{{}##}$\hfil\crcr#1\crcr}}\,}

\def\VEV#1{\left\langle #1\right\rangle}        

\begin{document}

\thispagestyle{empty}

{\hbox to\hsize{
\vbox{\noindent October 2017 \hfill IPMU17-0130 }}}
{\hbox to\hsize{
\vbox{\noindent   \hfill }}}

\noindent
\vskip2.0cm
\begin{center}

{\large\bf Inflation from higher dimensions}
\vglue.3in

Hiroshi Nakada~${}^{a}$ and Sergei V. Ketov~${}^{a,b,c}$ 
\vglue.1in

${}^a$~Department of Physics, Tokyo Metropolitan University, \\
Minami-ohsawa 1-1, Hachioji-shi, Tokyo 192-0397, Japan \\
${}^b$~Institute of Physics and Technology, Tomsk Polytechnic University,\\
30 Lenin Avenue, Tomsk 634050, Russian Federation \\
${}^c$~Kavli Institute for the Physics and Mathematics of the Universe (IPMU),
\\The University of Tokyo, Chiba 277-8568, Japan \\

\vglue.1in
nakada-hiroshi1@ed.tmu.ac.jp, ketov@tmu.ac.jp
\end{center}

\vglue.3in

\begin{center}
{\Large\bf Abstract} 
\end{center}
\vglue.1in

\noindent  We derive the scalar potential in four spacetime dimensions from an eight-dimensional $(R+\gamma R^4-2\Lambda-F_4^2)$ gravity model in the presence of the 4-form $F_4$, with the (modified gravity) coupling constant $\gamma$ and the cosmological constant  $\Lambda$, by using the flux compactification of four extra dimensions on a 4-sphere with the warp factor. The scalar potential depends upon two scalar fields: the scalaron and the 4-sphere volume modulus. We demonstrate that it gives rise to a viable description of cosmological inflation in the early Universe, with the scalaron playing the role of inflaton and  the volume modulus to be (almost) stabilized at its minimum. We also speculate about a possibility of embedding our model in eight dimensions into a modified eight-dimensional supergavity that, in its turn, arises from a modified eleven-dimensional supergravity.

\newpage

\section{Introduction}

Extra dimensions appear in Kaluza-Klein (KK) field theory and gravity, supersymmetry and supergravity, string theory and brane world, mainly in the context of unification of particles and fields. It is, therefore, natural to study multi-dimensional cosmological models also, and relate them to the observed Universe. However, the progress in this direction was limited in the literature, because the observed Universe is four-dimensional, so that any multi-dimensional cosmological model has to end up with the effective four-dimensional theory that fits the FLRW framework and is consistent with observations. Extra dimensions unavoidably   lead to extra scalar fields (called {\it moduli}) that must be stabilized.  In addition, the mass hierarchy has to be satisfied as follows:
\begin{equation} \label{hier}
M_{\rm inf.} \ll M_{\rm KK} \ll M_{\rm Pl}~~.
\end{equation} 
Moreover, extra dimensions usually open a lot of possibilities that should be constrained both theoretically and experimentally. This would imply interesting relations between the four-dimensional cosmological quantities and their higher dimensional counterparts, and offer a possible multi-dimensional origin of our Universe.

One of the well studied approaches in this direction is based on the modified $f(R)$ gravity actions in higher dimensions with a warped product geometry, where $R$ stands for the Ricci
scalar in $D>4$ spacetime dimensions.  However, as was demonstrated in
Refs.~\cite{zhuk1,zhuk2,zhuk3,zhuk4}, the higher-dimensional $(R+\gamma R^n-2\Lambda)$ gravity models together with their spontaneous compactification to four dimensions  {\it do not} lead to a successful phenomenological description of dark energy, because of a necessarily negative (induced) cosmological constant in four dimensions. These models also fail to describe the early universe inflation because of low values of the scalar index $n_s$ and the e-foldings number $N_e$. Adding extra (matter) $p$-form fields with a Freund-Rubin-like compactification ansatz \cite{frr} can stabilize extra dimensions for a certain range of parameters \cite{zhuk2}, but still does not lead to a successful phenomenology. In particular, the four-dimensional inflationary models based on the  compactified $(R+\gamma R^n-2\Lambda)$ gravity in dimensions $D<8$ were found to be not viable \cite{zhuk4}.  It raises a question about whether this situation can be improved by changing or relaxing some of the assumptions used in Refs.~\cite{zhuk1,zhuk2,zhuk3,zhuk4}. It is also desirable to get the constraints restricitng the values of a higher dimension $D>4$, the power $n$ of the scalar curvature $R$ in the modified gravity action, the value of the higher-dimensional cosmological constant $\L$, and the rank $p$ of a $p$-form field, if any.

In our paper \cite{kn2}, we proposed a derivation of the viable inflaton scalar potential from the 
higher $(D)$ dimensional  $(R+\gamma R^n-2\Lambda)$ gravity, by giving up the condition of
spontaneous compactificaton of extra dimensions and ignoring the moduli, i.e. just assuming that
the compactification happened before inflation and it can be made spontaneous by adding some more fields.  As a result, the inflaton scalar potential in four spacetime dimensions turns out to be dependent upon the parameters $(\gamma,\Lambda ,D,n)$, while the viable
inflationary phenomenology requires
\begin{equation} \label{nD}
n=D/2~~,
\end{equation}
with the dimension $D$ being a multiple of four. The condition (\ref{nD}) arises by demanding the existence of a {\it plateau} with a positive height for the inflationary scalar potential, as
is apparently favoured by the Planck mission observations 
\cite{Ade:2015xua,Ade:2015lrj,Array:2015xqh}, and is the case in the famous Starobinsky inflationary model \cite{Starobinsky:1980te}, though is in contrast to 
Refs.~\cite{zhuk1,zhuk2,zhuk3,zhuk4} where the scalar potential was demanded to vanish before the onset of inflation. Our results  were significantly enhanced in Ref.~\cite{nad} where 
a spontaneous compactification and stabilization of the volume of extra dimensions was achieved
by adding a single $(p-1)$-form gauge field having a non-vanishing flux in compact dimensions and  obeying the condition
\begin{equation} \label{np}
p=n~.
\end{equation}

In this paper we extend this analysis in the first relevant higher dimension $D=8$, and  consider an embedding of the $D=8$ modified gravity model into a (modified) $D=8$ supergravity. 

Our paper is organized as follows. Our modified gravity model in $D=8$ is formulated in Sec.~2.
Also in Sec.~2 we consider the Freund-Rubin-type compactification of our model on a 4-sphere down to four spacetime dimensions, derive the scalar potential, and stabilize the volume modulus of  the compact dimensions described by the 4-sphere. In Sec.~3 we apply our model to a description of cosmological inflation in the early Universe. In Sec.~4 we speculate about a possible embedding of our model into a modified $D=8$ supergravity. Sec.~5 is our conclusion. We collect all technical details into four appendices: appendix A is devoted to the Legendre-Weyl transform of the modified gravity model to the Einstein frame in 8 dimensions; appendix B describes the Freund-Rubin-type compactification of the transformed action to 4 dimensions on a 4-sphere, it includes a derivation of the two-field scalar potential in four dimensions; appendix C is devoted to a detailed study of the scalar potential in 4 dimensions; appendix D is devoted to a (partial) derivation of the (modified and gauged) $D=8$ supergravity from a modified $D=11$ supergravity by compactifying the latter on a 3-sphere.

\section{The $D=8$ model and its $D=4$ compactification}

The $f(R)$ gravity in {\it four} spacetime dimensions is the standard theoretical approach in modern cosmology, capable of describing both cosmological inflation in the early Universe and dark energy in the present Universe --- see e.g., the reviews \cite{svrev,tsu,clrev,myrev,starp} and references therein. The
basic idea is to replace the scalar curvature $R$ in the Einstein-Hilbert (EH) gravity action by a function $f(R)$ obeying certain physical requirements in the relevant range of its argument $R$, such as the absence of ghosts and tachyons, in order to fit the (observed) accelerating Universe.

The distinguished property of $f(R)$ gravity theories is their classical equivalence (duality) to the 
scalar-tensor gravity theories \cite{fmbook}, which is known for the long time --- see e.g., 
Refs.~\cite{wag,bick,whitt,jk,bc,maeda,kkw1,kw1,kw2}. The existence of this (Legendre-Weyl) transformation relating these apparently different gravity theories is guaranteed by the physical conditions on the $f(R)$-function, namely, positivity of its first and second derivatives (in the proper notation, and in the relevant range of the scalar curvature values). 

The simplest and, perhaps, most famous $f(R)$ gravity model of Starobinsky \cite{Starobinsky:1980te}
is defined by the action~\footnote{We use the natural units $\hbar=c=1$ with the reduced Planck mass $M_{\rm Pl}=1$, and the $D=4$ spacetime signature $(-,+,+,+)$.}
 \be \label{star}
S_{\rm Starobinsky} = \fracmm{1}{2} \int \mathrm{d}^4x\sqrt{-g} \left[ R +\fracmm{1}{6M^2}R^2\right]~.
\ee
The Starobinsky model is known as an excellent model of inflation, in very good agreement with the Planck data \cite{Ade:2015xua,Ade:2015lrj,Array:2015xqh}. On the one hand, {\it any} viable inflationary model with $f(R)=R+\tilde{f}(R)$ gravity must be close to the Starobinsky model (\ref{star}) in the sense of having 
 $\tilde{f}(R)=R^2 A(R)$ with a slowly-varying function  $A(R)$. The Starobinsky model is also known as an {\it attractor} for inflation \cite{gklr}. On the other hand, {\it any} $(R+\gamma R^n)$ gravity model in $D=4$ 
 with an integer power $n$ higher than two is {\it not} viable for 
 inflation \cite{kkw2}.~\footnote{Having $n$ to be non-integer and close to 2 is still possible for inflation  in $D=4$ \cite{moto}.} 

The only real parameter $M$ of the Starobinsky model can be identified with the inflaton mass, whose value is fixed by the observational Cosmic  Microwave Background (CMB) data as $M=(3.0 \times10^{-6})(\frac{50}{N_e})$ where $N_e$ is the e-foldings number.  The corresponding scalar potential of the (canonically normalized) inflaton field $\phi$ (dubbed {\it scalaron} in the given context) in the dual (scalar-tensor gravity) picture reads
\cite{jdb}
\begin{equation} \label{starpot}
V(\phi) = \fracmm{3}{4} M^2\left( 1- e^{-\sqrt{\frac{2}{3}}\phi }\right)^2~.
\end{equation}
This scalar potential is bounded from below, has a Minkowski vacuum and a {\it  plateau} of a positive height for slow roll inflation. During the inflation the scalar potential (\ref{starpot}) is simplified to
\begin{equation} \label{starpot2}
V(\phi) \approx V_0\left( 1- 2e^{-\alpha_s\phi }\right)~,
\end{equation}
where we have kept only the leading (exponentially small) correction to the emergent cosmological constant $V_0=\frac{3}{4} M^2$, and have introduced the notation $ \alpha_s=\sqrt{\frac{2}{3}}$~. 

It is the demand of having a plateau for the scalar potential in higher $D$ dimensions that results in
the condition (\ref{nD}) \cite{bc,kn2}. But it is still insufficient for moduli stabilization that requires
at least one $p$-form field obeying the condition (\ref{np}) \cite{nad}. 

Therefore, our minimal model in $D=8$ is defined by the action
\begin{equation} \lb{act1}
S= \fracmm{M_8^6}{2} \int d^8X\sqrt{-g_8} \left[
R_8+\gamma_8 R_8^4-2\Lambda_8-g^{A_1B_1}g^{A_2B_2}g^{A_3B_3} g^{A_4B_4}F_{A_1A_2A_3A_4}F_{B_1B_2B_3B_4}\right]. 
\end{equation}
It depends upon two fields, a metric $g_{AB}$ and a 3-form gauge potential $A_{ABC}$, whose
field strength 4-form is $F=dA$, and has three parameters: the gravitational mass scale $M_8$, 
 the (modified gravity) coupling constant $\gamma_8>0$ and the cosmological constant $\Lambda_8>0$, 
 all in $D=8$ dimensions --- see Appendix A for more details.

Applying the Legendre-Weyl transform to the action (\ref{act1}) in $D=8$ results in the dual (classically equivalent) action (see Appendix A for its derivation)
\begin{equation} \lb{act2}
\eqalign{
S_{\rm dual} = & \fracmm{M_8^6}{2} \int d^8X\sqrt{-\tilde{g}_8} \left[
\tilde{R}_8-42\tilde{g}^{AB}\partial_Af\partial_Bf  -M^2_8\tilde{V}(f) \right. \cr
 & \left. -\tilde{g}^{A_1B_1}\tilde{g}^{A_2B_2}\tilde{g}^{A_3B_3}\tilde{g}^{A_4B_4}F_{A_1A_2A_3A_4}F_{B_1B_2B_3B_4}\right]~~,\cr}
\end{equation}
depending upon three fields, the Weyl transformed (new) metric $\tilde{g}_{AB}$, the 4-form $F=dA$,
and the real scalaron  $f(X)$ having the scalar potential
\begin{equation} \lb{p1}
\tilde{V}(f)=  a^{-2}(1-e^{-6f})^{\frac{4}{3}}+2e^{-8f}\tilde{\Lambda}_8~~,
\end{equation}
in terms of the (dimensionless) coupling constants
\begin{equation} \lb{rdf}
\gamma_8 \equiv M_8^{-6}\tilde{\gamma}_8~~, \quad \Lambda_8\equiv M_8^{2}\tilde{\Lambda}_8~~,
\quad \fracmm{3}{4} \left(\fracmm{1}{4\tilde{\gamma}_8}\right)^{\frac{1}{3}}\equiv a^{-2}~~.
\end{equation}
The dual action ({\ref{act2}) has the standard form of Einstein's gravity coupled to the matter fields $(f,A)$ and
having the scalar potential (\ref{p1}) in $D=8$. This scalar potential has a plateau of the positive height $a^{-2}$ for large positive values of $f$.

Let us consider a compactification of the $D=8$ theory (\ref{act2}) on a 4-sphere $S^4$ with the warp factor  $\chi$, down to {\it four} spacetime dimensions, i.e. in a curved spacetime with the local structure
\begin{equation} \label{compactify}
{\cal M}_8 = {\cal M}_4\times e^{2\chi}S^4~~.
\end{equation}
The 8-dimensional coordinates $(X^A)$ can then be decomposed into the 4-dimensional spacetime coordinates $(x^{\alpha})$ with $\a=0,1,2,3$, and the coordinates $(y^a)$ of four compact dimensions of
$S^4$, with  $a,b,\ldots=1,2,3,4$. The compactification ansatz reads
\begin{equation} \lb{cansatz}
ds^2_8=\tilde{g}_{AB}dX^AdX^B=g_{\alpha\beta}dx^{\alpha}dx^{\beta}+e^{2\chi}g_{ab}dy^ady^b~~,
\end{equation} 
where $g_{\alpha\beta}=g_{\alpha\beta}(x)$, $g_{ab}=g_{ab}(y)$ and $\chi=\chi(x)$.

This compactification results in the following $D=4$ action (see Appendix B for its derivation): 
\begin{align} \lb{4tot}
S_4 [\hat{g}_{AB},\c, f]
&=\fracmm{M_{\rm Pl}^2}{2}\int d^4x\sqrt{-g}\left[ \hat{R}-12\hat{g}^{\alpha\beta}\partial_{\alpha}\chi\partial_{\beta}\chi \right. \nonumber\\
  &\left. - 42\hat{g}^{\alpha\beta}\partial_{\alpha}f\partial_{\beta}f
 - M_{\rm Pl}^2\left( e^{-4\chi}\tilde{V}(f)-2e^{-6\chi}-e^{-12\chi}F^2\right) \right]~~,
\end{align}
of three fields: a metric $\hat{g}_{\alpha\beta}(x)$, the scalaron $f(x)$ and the $S^4$ (volume) modulus 
$\chi(x)$, with the scalar potential depending upon the parameters $(a,\tilde{\Lambda}_8)$ {\it and}
 the 4-form gauge field strength {\it flux\/} $F$ defined by  the integration
 \begin{equation} \lb{fpar}
\int d^4y\sqrt{g_y}\,g^{a_1b_1}\cdots g^{a_4b_4}F_{a_1...a_4}F_{b_1...b_4} = M_8^{-2}F^2
\end{equation}
over the $S^4$. The full two-scalar potential in $D=4$ thus reads 
\begin{equation} \lb{2p}
M_{\rm Pl}^{-4}V(\chi,f)=\left[ a^{-2}(1-e^{-6f})^{\frac{4}{3}}+2\tilde{\Lambda}_8 e^{-8f}\right]e^{-4\chi}-2e^{-6\chi}+F^2e^{-12\chi}~~.
\end{equation}
We have restored the reduced Planck scale $M_{\rm Pl}$ in Eqs.~(\ref{4tot}) and (\ref{2p}) for reader's convenience.

The scalar potential (\ref{2p}}) is investigated in detail in Appendix C. It has a stable Minkowski vacuum
and a plateau with a positive height provided that
\begin{equation} \lb{Frange}
1<F^2/(16\tilde{\g}_8)\equiv 1+\d<(\frac{3}{2})^3~~,
\end{equation}
where the inequality on the right hand side is also needed to ensure a positive mass squared of the modulus $\chi$ at the onset of inflation --- see Eq.~(\ref{posh}).

For generic values of $\delta$ in Eq.~(\ref{Frange}) one gets a two-field inflation. However, the modulus
$\chi$ is strongly stabilized when $\d \ll 1$ that implies only a small shift of the minimum of $\chi$ during
inflation, from  $\c_c$ to $\c_0$, as
\be \lb{dds} 
 0<\c_c-\c_0 \approx \fracmm{1}{12}\d \ll 1~~,
\ee
and results in a {\it single-field inflation} driven by the inflaton (scalaron) $f$ in $D=4$.

The physical {\it hierarchy} of scales in Eq.~(\ref{hier}) can be satisfied provided that
\begin{equation} \lb{Fr}
F^2\gg 1~~,
\end{equation}
where we have used the KK scale 
\begin{equation} \lb{KKs}
M_{\rm KK}\approx e^{-\c_0}M_{\rm Pl}
\end{equation}
with the warp factor due to the compactification ansatz (\ref{compactify}) and (\ref{cansatz}).

The mass condition  $m\low{\hat{f}_0} <  m\low{\hat{\c}_0}$ implies $F^2/\tilde{\g}_8 <72$ that is
already satisfied due to (\ref{Frange}). However, it is not possible to get a stronger condition
 $m\low{\hat{f}_0} \ll  m\low{\hat{\c}_0}$.

A profile of the scalar potential in $D=4$ is given in Fig.~1. It should be mentioned that the cosmological constant 
in $D=8$ is given by Eq.~(\ref{fl8}) that implies 
\begin{equation}\lb{cc8}
\tilde{\Lambda}_8 = \fracmm{\d^{-1/3}}{2a^2}~~,
\end{equation}
where we have used Eq.~(\ref{rdf}) also. In particular, it means that $\d$ cannot vanish.

\begin{figure}[htbp] 
\begin{center}
\includegraphics[clip,width=7cm,height=5cm]{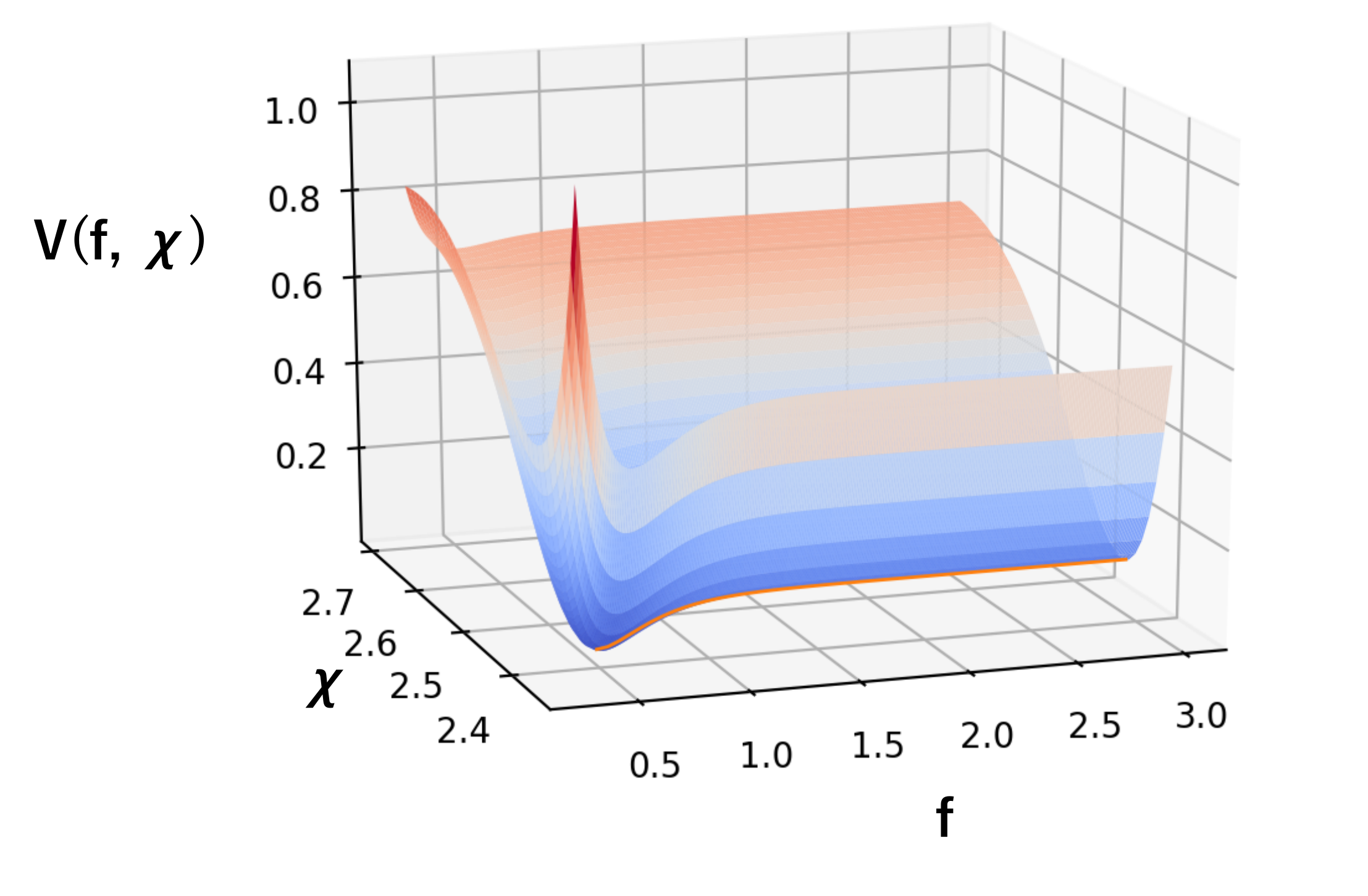}
\caption{The profile of the scalar potential (\ref{2p}) for the numerical input $F^2=10^6$, $\tilde{\g}_8= 6\cdot 10^4$ 
and $\tilde{\Lambda}_8\approx  0.0174$. The bottom line shows the inflationary trajectory.}
\label{fig:one}
\end{center}
\end{figure}

\section{Towards a supergravity embedding of our model}

In this Section we explore a possibility of embedding our 8-dimensional model (\ref{act1}) into a $D=8$ supergravity. First, supergravity may be the natural origin of the $p$-form field because higher-dimensional supergravities usually include such fields. Second, the supergravity extensions of modified gravity certainly exist  in $D=4$ \cite{myrev,kter}, and they should also exist in higher dimensions $D\leq 11$. 

Unfortunately, to the best of our knowledge, no fully supersymmetric extension of any $(R+R^4)$ gravity
in higher $(8\leq D \leq 11)$  dimensions was ever derived, so that our investigation in this Section cannot be fully consistent and compelling, unlike that in the previous Sections. Moreover, any standard (two-derivative) supergravity theory does not allow a positive cosmological constant in its action (it would break supersymmetry), so that the origin of the cosmological consant in $D=8$ can only be either due to a spontaneous supersymmetry breaking or/and some nonperturbative effects. So, this Section ends up with a conjecture.

A good starting point of this investigation is the maximally supersymmetric $D=11$ supergravity, because
of its uniqueness. It can be modified by the quartic scalar curvature term and then compactified down to
$D=8$ on a compact manifold (3-sphere $S^3$) --- see Appendix D for details. Moreover, the $SO(3)$  non-abelian isometries of the $S^3$ can be gauged, thus producing the non-abelian gauge fields and a scalar potential in $D=8$. Taken together, it leads to the bosonic part of the (modified and gauged) $D=8$ supergravity action, having the form (\ref{8saction}),
\begin{gather} 
S_8=\int d^8x\fracmm{e}{2\kappa^2}[R+\tilde{\gamma}e^{2\kappa\phi}R^4-\kappa^2e^{2\kappa\phi}F_{\mu\nu}^\alpha F^{\mu\nu}_\alpha-2\kappa^2\partial_\mu\phi\partial^\mu\phi-V(T)-P_{\mu ij}P^{\mu ij}\nonumber\\
-\fracmm{1}{2}\kappa^2e^{-4\kappa\phi}\partial_\mu B\partial^\mu B-\fracmm{\tilde{\kappa}^2}{12}e^{2\kappa\phi}G_{\mu\nu\rho\sigma}G^{\mu\nu\rho\sigma}-\fracmm{\kappa^3}{432}e^{-1}\varepsilon^{\mu_1...\mu_8}G_{\mu_1...\mu_4}G_{\mu_5...\mu_8}B]+...~, \label{finale}
\end{gather}
in terms of the following $D=8$ fields: a metric $g_{\m\n}$, dilaton $\f$, the $SO(3)$ gauge field strength
$F^{\a}_{\m\n}$, the vector fields  $P_{\mu ij}$, the 4-form gauge field strength  $G_{\mu\nu\rho\sigma}$ and $(5+1)$ scalars $(T,B)$ whose scalar potential is
\begin{equation} \lb{spt8}
V(T)=\fracmm{g^2}{4\kappa^2}e^{-2\kappa\phi}(T_{ij}T^{ij}-\frac{1}{2}T^2)~.
\end{equation}
The supergravity (\ref{finale})  has the required {\it quartic} scalar curvature term and the gauge 3-form kinetic term given by the gauge field strength {\it 4-form} squared, while the abelian vector fields  $P_{\mu ij}$ are merely the spectators here. Hence, (\ref{finale}) could be the supersymmetric extension of our action (\ref{act1}) provided that (i) the dilaton $\f$ is stabiized, and (ii) a positive cosmological constant is generated. One usually assumes in the literature that the dilaton potential is generated by quantum non-perturbative corrections beyond the supergravity level. And the cosmological constant may be generated by the non-perturbative vacuum expectation value
\begin{equation} \label{vevF}
\VEV{\k^2e^{2\kappa\phi}F_{\mu\nu}^\alpha F^{\mu\nu}_{\a}}=2\Lambda_8~~.
\end{equation}
Unfortunately, we do not have means to compute the dilaton vacuum expectation value and the 
gluon condensate (\ref{vevF}) in $D=8$. 

\section{Inflation}

Once the modulus $\chi$ is strongly stabilized (Sec.~2), the inflaton potential (\ref{2p}) takes the form 
($M_{\rm Pl}=1$) 
\begin{equation} \lb{effsp}
e^{4\c_0}a^2V(f)=\left(1-e^{-6f}\right)^{\frac{4}{3}}+\lambda e^{-8f} - \lambda(1+\lambda^3)^{-\frac{1}{3}}~~.
\end{equation}
with $\l=2a^2\tilde{\Lambda}_8=\d^{-1/3}$. This potential has the absolute minimum at 
\begin{equation} \lb{min8}
f_0=\frac{1}{6}\ln\left(1+\lambda^3\right)~~,
\end{equation}
where it vanishes in the Minkowski vacuum. A profile of the scalar potential (\ref{effsp}) is given in Fig.~2.

\begin{figure}[htbp] 
\begin{center}
\includegraphics[clip,width=7cm,height=5cm]{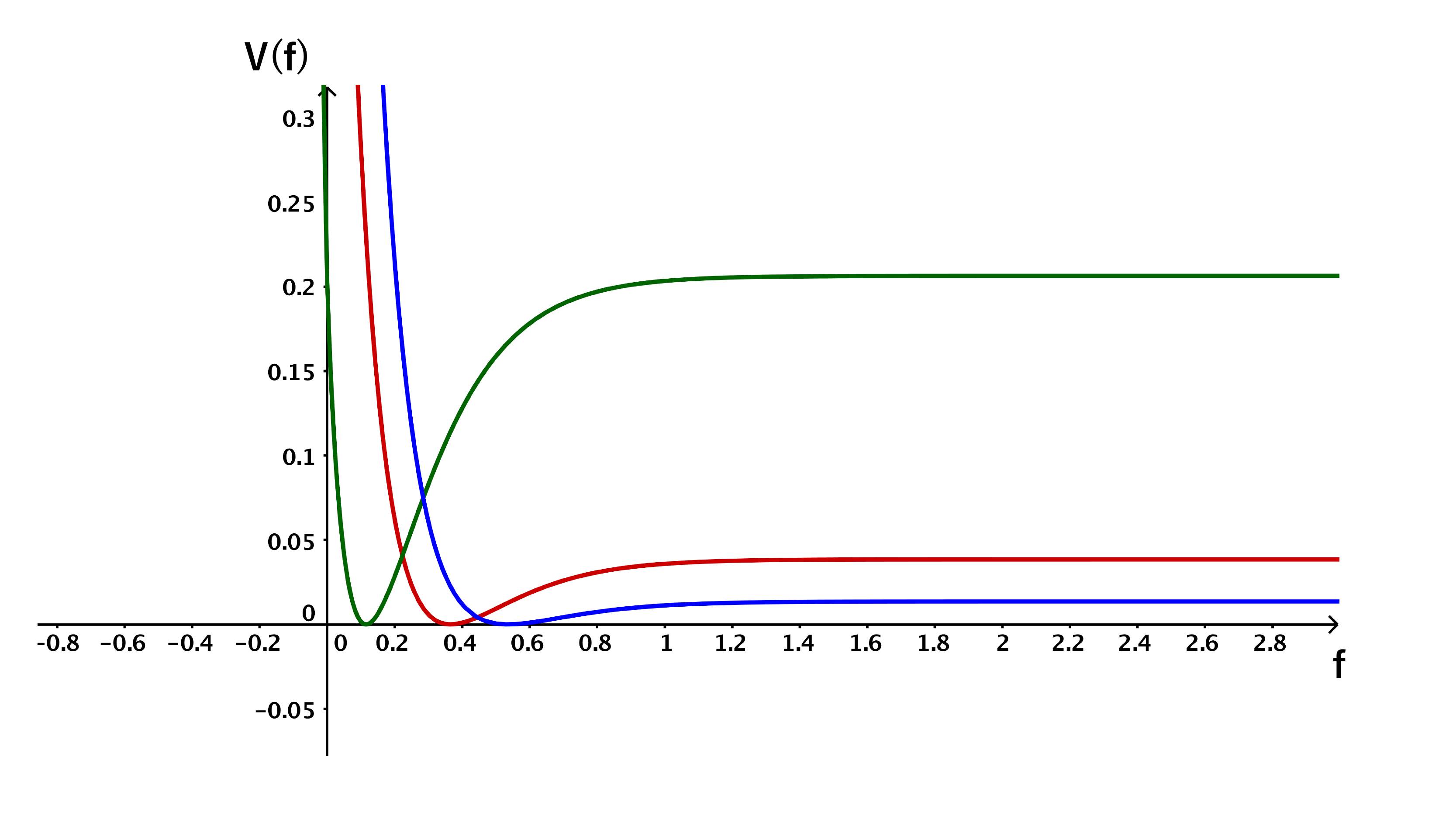}
\caption{The profile of the scalar potential (\ref{effsp}) for $\lambda=1$ (green), $\lambda=2$ (red) and
$\lambda=2.88$ (blue).}
\label{fig:two}
\end{center}
\end{figure}

During inflationary slow roll along the plateau, the scalar potential  (\ref{effsp})  can be approximated as
\begin{equation} \label{pclass}
V(\phi) = V_0 - V_1e^{-\alpha\phi}~~,
\end{equation}
with
\begin{equation} \label{aval}
\alpha = \sqrt{\fracmm{6}{7}}~~.
\end{equation}
This value  of $\alpha$ determines the key observational parameter $r$ related to primordial gravity waves and known as the tensor-to-scalar ratio, 
\begin{equation} \lb{rbound}
r =\fracmm{8}{\alpha^2N_e^2}=\fracmm{28}{3N_e^2}~~.
\end{equation}
The Planck data \cite{Ade:2015lrj} sets the upper bound on $r$ (with 95\% of CL) as $r<0.08$. It implies 
\begin{equation}\lb{abound}
\alpha >\fracmm{10}{N_e}= 0.2 \left(\fracmm{50}{N_e}\right)~,
\end{equation}
while our result (\ref{aval}) is clearly  above this bound.

As regards the other CMB spectral tilts (the inflationary observables), the scalar spectral index $n_s$ 
and its running $dn_s/d{\ln}k$, their values derived from the potential  (\ref{pclass}) are
\begin{equation} \label{otilts}
n_s\approx 1 - \fracmm{2}{N_e} \qquad {\rm and} \qquad \fracmm{dn_s}{d \ln k} \approx -\fracmm{(1-n_s)^2}{2}\approx -\fracmm{2}{N_e^2}~~,
\end{equation}
i.e. they are the same as in the Starobinsky model (\ref{star}) and (\ref{starpot}).

The microscopic parameters of our model can be easily tuned to get the {\it same} inflaton mass $M$, so that
our effective inflationary model obtained from the higher ($D=8$) dimensions is almost indistinguishable from 
the orginal Starobinsky model having $\a_s=\sqrt{2/3}$. 

 When a conventional matter action is added to our gravity model, Weyl rescalings of the metric result in the {\it universal} couplings (via the GR covariant derivatives) of inflaton $f$ to all matter fields with powers of $\exp{(-\alpha\kappa_4 f)}$. The value (\ref{aval}) of $\alpha$ derived from $D=8$ is only slightly different from the Starobinsky value $\a_s=\sqrt{2/3}$, while all the matter couplings to the scalaron are suppressed by the Planck mass. Therefore, the impact of higher dimensions on the inflationary observables and reheating is negligible in
 our approach.

\section{Conclusion}

Our main results are summarized in the Abstract. 

We used the Starobinsky inflationary model of the $(R+R^2)$ gravity (\ref{star}) in four dimensions as the prototype for deriving the new inflationary models from higher dimensions. Among the advantages of this approach are (i) its geometrical nature, as only gravitational interactions are used, (ii) consistency with the current astronomical observations of CMB, (iii)
the clear physical nature of inflaton (scalaron) as the spin-0 part of metric. In this paper we focused on $D=8$ dimensions only. In our scenario, the Universe was born multi-dimensional, and then four spacetime dimensions became infinite, while the others curled up by some unknown mechanism before inflation. The inflation happened after the compactification and the moduli stabilization. 

In higher-dimensions, it turned out to be necessary to include a cosmological constant
and a gauge (form) field, with the strong conditions on the higher dimension, the power $n$ of the scalar curvature  
and the rank of the form, see Eqs.~(\ref{nD}) and (\ref{np}). The moduli stabilization and the scale hierarchy are also possible to achieve, while both are non-trivial in the present context. It may also be possible to embed our $D=8$ modified gravity model into the modified $D=8$ supergravity and, ultimately, into the modified $D=11$ supergravity.

As regards the observational predictions of our model, it leads to the certain value (\ref{rbound}) of the CMB tensor-to-scalar ratio that is, however, only slightly different from that of the original Starobinsky model.

Our results may be used for studying inflation and moduli stabilization in more general frameworks, such as unification of
fields and forces, KK theories of gravity, supergravity and superstrings, and braneworld.~\footnote{In particular, as was found in \cite{nk1}, the modified $(R+R^2)$ gravity in the Randall-Sundrum (RSII) braneworld \cite{rs} does {\it not} destabilize the 
famous Randall-Sundrum solution to the hierarchy problem in particle physics.}
\vglue.2in

\section*{Acknowledgements}

S.V.K. was supported by a Grant-in-Aid of the Japanese Society for Promotion of Science (JSPS) under No.~26400252, the Competitiveness Enhancement Program of Tomsk Polytechnic University in Russia, and the World Premier International Research Centre Initiative (WPI Initiative), MEXT, Japan. S.V.K. is grateful to A. Zhuk and Y. Aldabergenov for discussions and correspondence.
\vglue.2in

\section*{Appendix A: Legendre-Weyl transform in $D=8$}

We denote spacetime vector indices in eight dimensions by capital latin letters $A,B,\ldots=0,1,\ldots, 7$, and use the spacetime signature $(-,+,\ldots,+)$.

Let us begin with the following gravitational action in an 8-dimensional curved spacetime:
\begin{equation} \lb{action1}
 S_{8,{\rm grav.}}=\fracmm{1}{2\kappa_8^2}\int d^8X\sqrt{-g_8}(R_8+\gamma_8 R_8^4-2\Lambda_8)~~,
\end{equation}
where we have introduced the gravitational coupling constant $\kappa_8$ of (mass) dimension 
$(-3)$, the (modified gravity) coupling constant $\gamma_8>0$ of (mass) dimension $(-6)$, and 
and the cosmological constant $\Lambda_8>0$ of (mass) dimension $(+2)$, all in $8$ dimensions.~\footnote{The results of this Appendix are obtained by specifying the more general
results (for any $D$ and $n$) of Ref.~\cite{kn2} to $D=8$ and $n=4$.}

The action (\ref{action1}) can be rewritten to the form
\begin{equation}  \lb{action2}
S_{8,{\rm grav.}} =\fracmm{1}{2\kappa^2_8}\int d^8X\sqrt{-g_8}\left[(1+
 B)R_8 - \fracmm{3}{4}
 \left(\fracmm{B^4}{4\gamma_8}\right)^{\frac{1}{3}}  -2\Lambda_8\right]~~,
\end{equation}
where we have introduced the new scalar field $B$. The field $B$ enters the action 
(\ref{action2})  algebraically, while its "equation of motion" reads 
\begin{equation}  \lb{eom}
B=4\gamma_8 R^{3}_8~~.
\end{equation} 
Substituting it back into the action (\ref{action2})  yields the original action (\ref{action1}). Hence, the actions (\ref{action1}) and (\ref{action2}) are classically equivalent. 

In order to transform the action (\ref{action2}) to Einstein frame, we apply a Weyl transformation of the metric with the spacetime-dependent parameter $\Omega(X)$ in $8$ dimensions,
\begin{equation} \lb{weylt}
 g_{AB}=\Omega^{-2}\tilde{g}_{ AB},\ \ \ \sqrt{-g_8}=\Omega^{-8}\sqrt{-\tilde{g}_8}~~,
\end{equation}
where we have introduced the new spacetime metric $\tilde{g}_{ AB}$. As a result of this transformation, the (Ricci) scalar curvature gets transformed as
\begin{equation} \lb{weylr}
 R_8=\Omega^2[\tilde{R}_8+14\Tilde{\square}_8f-42\tilde{g}^{AB}f_{A}f_{B}]~~,
\end{equation}
where we have introduced the notation
\begin{equation}
 f=\ln \Omega~,\qquad f_{A}=\fracmm{\partial_A\Omega}{\Omega}~~,
\end{equation}
and the covariant wave operator $\Tilde{\square}_8=\Tilde{D}^A\Tilde{D}_A$ in $8$ dimensions.

The Weyl-transformed (and also equivalent by the field-redefinition (\ref{weylt})) action reads
\begin{align} \lb{action3}
 S_{8,{\rm grav.}}=&\fracmm{1}{2\kappa_8^2}\int d^8X\sqrt{-\tilde{g}_{ 8}}\Omega^{-8}\left[ (1+B)\Omega^2(\tilde{R}_8+14\Tilde{\square}_8f \right. \nonumber\\
 &\left. -42\tilde{g}^{AB}f_Af_B)- \fracmm{3}{4} \left(\fracmm{B^4}{4\gamma_8}\right)^{\frac{1}{3}} 
 -2\Lambda_8 \right]~.
\end{align}

Hence, the action in the Einstein frame is obtained by choosing the local parameter $\Omega$ as
\begin{equation} \lb{omegaf}
\Omega^{6}=e^{6f}=1+B~~.
\end{equation}
After ignoring the total derivative in the Lagrangian, it yields
\begin{align} \lb{action4}
 S_{8,{\rm grav.}}[\tilde{g}_{AB},f] =&\fracmm{1}{2\kappa^2_8} \int d^8X \sqrt{-\tilde{g}_{8}} \left[ \tilde{R}_8
 -42\tilde{g}^{AB}\partial_Af\partial_Bf \right. \nonumber\\
 - & \left. \fracmm{3}{4} \left(\fracmm{1}{4\gamma_8}\right)^{\frac{1}{3}}\left(1-e^{-6f}\right)^{4/3} -
 2e^{-8f}\Lambda_8 \right]~~.
\end{align}

Let us redefine the coupling constants in $8$ dimensions as 
\begin{equation} \lb{ccrdf}
\kappa_8\equiv M_8^{-3}~~,\quad \gamma_8 \equiv M_8^{-6}\tilde{\gamma}_8~~,
\quad \Lambda_8\equiv M_8^{2}\tilde{\Lambda}_8~~,\quad 
\fracmm{3}{4} \left(\fracmm{1}{4\tilde{\gamma}_8}\right)^{\frac{1}{3}}\equiv a^{-2}~~,
\end{equation}
in terms of the new (mass) parameter $M_8>0$ of dimension $(+1)$, and the dimensionless
parameters $\tilde{\Lambda}_8$ and $a>0$. Then the action (\ref{action4}) takes the form
\begin{equation} \lb{action5}
  S_{8,{\rm grav.}}[\tilde{g}_{AB},f] =\fracmm{M_8^6}{2} \int
  d^8X\sqrt{-\tilde{g}_{8}}\left[ \tilde{R}_8-42\tilde{g}^{AB}\partial_Af\partial_Bf
  -M^2_8\tilde{V}(f)\right]
\end{equation}
with the (dimensionless) scalar potential
\begin{equation} \lb{pot1}
\tilde{V}(f)=  a^{-2}(1-e^{-6f})^{\frac{4}{3}}+2e^{-8f}\tilde{\Lambda}_8~~.
\end{equation}

Given an 8-dimensional action of the 4-form $F$ (the totally antisymmetric gauge field strength $F=dA$ of a gauge 3-form potential $A$) in the form  
\begin{equation} \lb{4faction1}
 S_{8} [g_{AB},F_4] =- \fracmm{M_8^6}{2}\int d^8X\sqrt{-g_8}\,g^{A_1B_1}g^{A_2B_2}g^{A_3B_3} g^{A_4B_4}F_{A_1A_2A_3A_4}F_{B_1B_2B_3B_4}~~,
\end{equation}
the $F$ has (mass) dimension $(+1)$, and the $A$ is dimensionless.

Under the Weyl transform (\ref{weylt}), the $\Omega$-factors are {\it cancelled} against
each other, so that the action (\ref{4faction1}) remains unchanged,
\begin{equation} \lb{4faction2}
 S_{8} [\tilde{g}_{AB},F_4] =-\fracmm{M_8^6}{2}\int d^8X\sqrt{-\tilde{g}_8}\,\tilde{g}^{A_1B_1}\cdots  \tilde{g}^{A_4B_4}F_{A_1...A_4}F_{B_1...B_4}~~.
\end{equation}       

The action of our model in 8 dimensions is defined by
\begin{equation} \lb{action6}
S_8 [\tilde{g}_{AB},f,F_4] = S_{8,{\rm grav.}}[\tilde{g}_{AB},f] +  S_{8} [\tilde{g}_{AB},F_4] ~~.
\end{equation}   

\section*{Appendix B: Freund-Rubin-type compactificaton}

In this Appendix we consider the compactification of the theory (\ref{action6}) on a 4-sphere $S^4$ with the warp factor $\chi$, down to 4 spacetime dimensions. We separate the 8-dimensional coordinates $(X^A)$ into the 4-dimensional spacetime coordinates $(x^{\alpha})$
with $\a=0,1,2,3$, and the coordinates $(y^a)$ of four extra (compact) dimensions with 
$a,b,\ldots=1,2,3,4$.~\footnote{Our results in this Appendix fully agree with Ref.~\cite{nad}.}

We use the standard compactification ansatz
\begin{equation} \lb{mansatz}
ds^2_8=\tilde{g}_{AB}dX^AdX^B=g_{\alpha\beta}dx^{\alpha}dx^{\beta}+e^{2\chi}g_{ab}dy^ady^b~~,
\end{equation} 
where $g_{\alpha\beta}=g_{\alpha\beta}(x)$, $g_{ab}=g_{ab}(y)$ and $\chi=\chi(x)$, with the
normalization
\begin{equation} \lb{norm4s}
  \int d^4y\sqrt{g_y}=M_8^{-4}~~.
\end{equation}  
Taking into account the  $S^4$ Euler number equal to 2, yields
 \begin{equation} \lb{eulers}
 \int d^4y\sqrt{g_y}R_y=2M_8^{-2}~~,
\end{equation}
where $R_y$ is the scalar curvature of the sphere $S^4$. The decomposition (\ref{mansatz}) also implies 
\begin{equation} \lb{measured}
\sqrt{-\tilde{g}_8}=e^{4\chi}\sqrt{-g_4}\sqrt{g_{y}}
\end{equation}
and
\begin{equation} \lb{riccideco}
\tilde{R}_8 =R+e^{-2\chi}R_{y}-8e^{-\chi} \Tilde{\square}e^{\chi}
- 12 e^{-2\chi}g^{\alpha\beta} \partial_{\alpha}e^{\chi}\partial_{\beta}e^{\chi}~~,
\end{equation}
where we have introduced the Ricci scalar $R$ and the generally covariant wave operator 
$\Tilde{\square}=g^{\alpha\beta}\nabla_{\alpha}\nabla_{\beta}$ in four spacetime dimensions. 

The volume $\cv$ of four (compact) extra dimensions is given by 
\begin{equation} \lb{4smod}
\cv =   \int d^4y\sqrt{ \det(e^{2\chi} g_{ab})} = e^{4\chi}M_8^4~~,
\end{equation}
so that the warp factor $\chi$ can be identified with the volume modulus of the sphere $S^4$.

A substitution of  Eqs.~(\ref{mansatz}), (\ref{measured}) and (\ref{riccideco}) into the action (\ref{action5}), and an integration over the compact dimensions by using Eqs.~(\ref{norm4s}) and
(\ref{eulers}), lead to the action
\begin{align}\lb{4action1}
 S_4[g_{\a\b},f,\c] 
 &=\fracmm{M_8^2e^{4\chi_0}}{2}\int
 d^4x\sqrt{-g}\left(\fracmm{e^{\chi}}{e^{\chi_0}}\right)^4\left[ R+2M_8^2e^{-2\chi}\right. \nonumber \\
 &\left. +12g^{\alpha\beta}\partial_{\alpha}\chi\partial_{\beta}\chi-42g^{\alpha\beta}\partial_{\alpha}f\partial_{\beta}f-M_8^2\tilde{V}(f)\right]~~,
\end{align}
where we have introduced the vacuum expectation value $\VEV{\c}_0=\c_0=const$. 

The action (\ref{4action1}) is still in a Jordan frame, so that the wrong sign of the kinetic term of the field $\c$ is not necessarily a problem. The Weyl transformation with the parameter 
$h(x)$ to the Einstein frame is given by
\begin{equation} \lb{weyl2a}
 g_{\alpha\beta}=e^{-2h}\hat{g}_{\alpha\beta},\quad
 h=2(\c-\c_0)~~.
\end{equation}
It implies 
\begin{equation} \lb{weyl2b}
 g^{\alpha\beta}=e^{2h}\hat{g}^{\alpha\beta},\quad \sqrt{-g}=e^{-4h}\sqrt{-\hat{g}}~,
\end{equation}
and
\begin{equation} \lb{weyl2c}
 R=e^{2h}\left[\hat{R}+6\hat{g}^{\alpha\beta}\nabla_{\alpha}\nabla_{\beta}h-6\hat{g}^{\alpha\beta}\partial_{\alpha}h\partial_{\beta}h\right]~~.
\end{equation}
Accordingly, the action (\ref{4action1}) gets transformed to 
 \begin{align} \lb{4action2}
  S_4 [\hat{g}_{\a\b},f,\c]  &=\fracmm{M_8^2e^{4\chi_0}}{2}\int
  d^4x\sqrt{-\hat{g}_4}\left\{ \hat{R}-12\hat{g}^{\alpha\beta}\partial_{\alpha}\c\partial_{\beta}\c
  \right. \nonumber\\
  &\left. -42\hat{g}^{\alpha\beta}\partial_{\alpha}f\partial_{\beta}f
  -\left(\fracmm{e^{\chi}}{e^{\chi_0}}\right)^{-4}M_8^2\left[ \tilde{V}(f)-2e^{-2\chi}\right]\right\}~,
 \end{align}
with the {\it physical} signs in front of all the kinetic terms. This also fixes the four-dimensional (reduced) Planck mass as
\begin{equation}\lb{planckm}
M_{\rm Pl}^2\equiv \k^{-2}= M_8^2e^{4\chi_0}~.
\end{equation}
Therefore, we have
\begin{align}  \lb{4action3}
 S_4  [\hat{g}_{\a\b},f,\c] 
 &= \fracmm{M^2_{\rm Pl}}{2}\int
 d^4x\sqrt{-\hat{g}}\left[ \hat{R}-12\hat{g}^{\alpha\beta}\partial_{\alpha}\chi\partial_{\beta}\chi
 \right. \nonumber \\
 -&\left. 42\hat{g}^{\alpha\beta}\partial_{\alpha}f\partial_{\beta}f
 -e^{-4\chi} M^2_{\rm Pl} \left( \tilde{V}(f)-2e^{-2\chi}\right)\right]~~.
\end{align}

Similarly, applying the compactification ansatz (\ref{mansatz}) to the 4-form action 
(\ref{4faction2}) in 8 dimensions yields
 \begin{equation} \lb{8actionF}
 S_{8,F} [\tilde{g}_{AB},F]=-\fracmm{M_8^6}{2}\int d^4x\sqrt{-g}\int d^4y\sqrt{g_y}\,e^{-4\chi}g^{a_1b_1}\cdots g^{a_4b_4}F_{a_1...a_4}F_{b_1...b_4}~~.
\end{equation}       
After defining the (dimensionless) {\it flux\/} parameter $F^2$ as 
\begin{equation} \lb{fluxp}
\int d^4y\sqrt{g_y}\,g^{a_1b_1}\cdots g^{a_4b_4}F_{a_1...a_4}F_{b_1...b_4} = M_8^{-2}F^2=const.~~,
\end{equation}
and using the Weyl transformation (\ref{weyl2a}), the action (\ref{8actionF}) reduces to
\begin{align} \lb{4actionF}
S_{4,F}[\hat{g}_{AB},\c]
&=-\fracmm{M_8^2e^{4\chi_0}}{2}\int d^4x\sqrt{-g}\left(\fracmm{e^{\chi}}{e^{\chi_0}}\right)^4e^{-8\chi}M_8^2F^2\nonumber \\
&=-\fracmm{M_8^2e^{4\chi_0}}{2}\int d^4x\sqrt{-\hat{g}}\,e^{-4h}\left(\fracmm{e^{\chi}}{e^{\chi_0}}\right)^4e^{-8\chi}M^2F^2\nonumber \\
&=-\fracmm{M^4_{\rm Pl}}{2}\int d^4x\sqrt{-\hat{g}}\,e^{-12\chi}F^2~~.
\end{align}

The total action in 4 dimensions is given by a sum of Eqs.~(\ref{4action3}) and (\ref{4actionF}),
\begin{align} \lb{4total}
S_4 [\hat{g}_{AB},\c, f]
&=\fracmm{M_{\rm Pl}^2}{2}\int d^4x\sqrt{-g}\left[ \hat{R}-12\hat{g}^{\alpha\beta}\partial_{\alpha}\chi\partial_{\beta}\chi \right. \nonumber\\
  &\left. - 42\hat{g}^{\alpha\beta}\partial_{\alpha}f\partial_{\beta}f
 - M_{\rm Pl}^2\left( e^{-4\chi}\tilde{V}(f)-2e^{-6\chi}-e^{-12\chi}F^2\right) \right]~~.
\end{align}

The canonical scalar fields $\hat{\chi}$ and $\hat{f}$ are thus given by
\begin{equation} \lb{norm2s}
\hat{\chi}=2\sqrt{3}M_{\rm Pl}\chi \quad {\rm and} \quad  \hat{f}=\sqrt{42}M_{\rm Pl}f~~,
\end{equation}
and the two-scalar potential in 4 dimensions reads 
\begin{equation} \lb{2pot}
M_{\rm Pl}^{-4}V(\chi,f)=\left[ a^{-2}(1-e^{-6f})^{\frac{4}{3}}+2\tilde{\Lambda}_8 e^{-8f}\right]e^{-4\chi}-2e^{-6\chi}+F^2e^{-12\chi}~~.
\end{equation}

\section*{Appendix C : study of the scalar potential}

In this Appendix we investigate the scalar potential (\ref{2pot}) in four dimensions. It depends upon
two fields, the inflaton $f$ and the modulus $\c$, and has three parameters $(a^{-2},F^2,\tilde{\L}_8)$ originating from eight dimensions (see appendices A and B).~\footnote{A partial analysis of generic potentials arising in the
same way from any dimension $D=2n$ was done in Ref.~\cite{nad}. We get more results for $D=8$.}

The potential (\ref{2pot}) has a Minkowski vacuum $(f_0,\c_0)$ defined by the equations
\begin{equation} \lb{mvac}
\left. \fracmm{\pa V}{\pa f} \right|_{f=f_0}=\left. \fracmm{\pa V}{\pa \c} \right|_{\c=\c_0}=\left.
V\right|_{f=f_0,~\c=\c_0}=0~~.
\end{equation}
The solution to these three equations is given by
\begin{equation} \lb{smvac}
e^{6f_0}= 1 + (2\tilde{\L}_8a^2)^3 \quad {\rm and} \quad  e^{6\c_0}= 2F^2~,
\end{equation}
together with a condition of the parameters,
\begin{equation} \lb{fl8}
\frac{2}{3}\tilde{\L}_8=\left( \fracmm{1}{16F^2 - 256 \tilde{\g}_8}\right)^{1/3}~~,
\end{equation}
where we have used the third relation (\ref{ccrdf}) between $\tilde{\g}_8$ and $a$.

The second derivatives of the scalar potential (\ref{2pot}) at the critical point (\ref{smvac}) determine the masses of the canonically normalized scalars  (\ref{norm2s}) as
\be \lb{massf}
m^2_{\hat{f}_0}=\left. \fracmm{\pa^2 V}{\pa f^2} \right|_{f=f_0}\fracmm{1}{42M^2_{\rm Pl}} =
\fracmm{M^2_{\rm Pl}}{56F^2}\left( \fracmm{F^2}{\tilde{\g}_8} - 16\right)~~, 
\ee
and
\be \lb{massh}
m^2_{\hat{\c}_0}=\left. \fracmm{\pa^2 V}{\pa \c^2} \right|_{\c=\c_0}\fracmm{1}{12M^2_{\rm Pl}} =
\fracmm{M^2_{\rm Pl}}{F^2}~~,
\ee
where we have used (\ref{fl8}) also. Equations (\ref{fl8}) and (\ref{massf}) imply the same condition
\be \lb{stab}
\fracmm{F^2}{\tilde{\g}_8}>16
\ee
for both the existence of a Minkowski vacuum and its stability, respectively.

At the onset of inflation ($f=+\infty$), the scalar potential of the modulus $\c$ is given by
\begin{equation} \lb{hpot}
M_{\rm Pl}^{-4}V(\chi)=a^{-2}e^{-4\chi}-2e^{-6\chi}+F^2e^{-12\chi}
\end{equation}
that only depends upon two (free) parameters $(a^{-2},F^2)$.

The critical points of the potential (\ref{hpot}) are determined by the condition
\begin{equation} \lb{hpotcr}
a^{-2}-3e^{2\chi_c}+3F^2e^{-8\chi_c}=0
\end{equation}
that has the form of the {\it depressed quartic} equation 
\begin{equation} \lb{dqua} 
z^4 + qz +r =0
\end{equation}
in terms of
\begin{equation} \lb{quadef}
z= e^{-2\c_c}~~,\qquad q=\fracmm{-1}{F^2}<0~,\qquad r = \fracmm{1}{3a^2F^2}>0~~.
\end{equation}
The {\it quartic} discriminant is given by 
\begin{equation} \lb{quadis}
\fracmm{\D_4}{27\cdot 256} = (r/3)^3 - (q/4)^4 ~~,
\end{equation}
while writing down an explicit solution to (\ref{dqua}) depends upon the sign of $\D_4$. 

By using the auxiliary (Ferrari's) resolvent cubic equation
\begin{equation} \lb{ferc}
m^3 - rm - q^2/8=0~~,
\end{equation}
we can factorize the left-hand-side of the quartic equation (\ref{dqua}) as
\begin{equation} \lb{quafact}
\left(z^2+m +\sqrt{2m} z - \fracmm{q}{2\sqrt{2m}}\right)\left(z^2+m -\sqrt{2m} z + \fracmm{q}{2\sqrt{2m}}\right)=0~.
\end{equation}
Because each term in the first factor is positive in our case, we get a {\it quadratic} 
equation from the vanishing second factor whose two roots are given by
\be \lb{qurt2}
z_{1,2} = \sqrt{\fracmm{m}{2}}\left[ 1 \pm \sqrt{-\fracmm{q}{m}-\sqrt{2m} }\right]~.
\ee
These two roots precisely correspond to the existence of a local (meta-stable) minimum and a local maximum of the potential (\ref{hpot}), with $-\infty<\c_{\rm min.}<\c_{\rm max.}<+\infty$.

The {\it cubic} discriminant $\D_3=4r^3-27(q^2/8)^2$ of the {\it depressed} cubic equation (\ref{ferc}) is
simply related to $\D_4$ as
\begin{equation} \lb{qiscrrel}
\fracmm{\D_3}{4\cdot 27} = (r/3)^3 - (q/4)^4 = \fracmm{\D_4}{27\cdot 256}~~~.
\end{equation}

When $\D_{3,4}\geq 0$, three real solutions to the cubic equation (\ref{ferc}) are given by the Viet\'e formula
\be \lb{viete}
m_{k}=2\sqrt{r/3}\cos\theta_k~~,\qquad k=0,1,2~,
\ee
where 
\be \lb{viete2}
\theta_k = \fracmm{1}{3}\arccos \left( \fracmm{3q^2}{16r}\sqrt{3/r}\right)-\fracmm{2\p k}{3}~,
\ee
while we should choose the highest (positive) root. The condition $\D_{3,4}\geq 0$ implies
\be \lb{dicsp}
\fracmm{F^2}{\tilde{\g}_8}\geq 27~~.
\ee

When $\D_{3,4}\leq 0$ or, equivalently, $F^2/\tilde{\g}_8\leq 27$, the angle (\ref{viete2}) does not
exist. Instead, we should use the Viet\'e's substitution in Ferrari's equation with
\be \lb{vietesub}
m = w +\fracmm{r}{3w}~,\qquad r>0~~,
\ee
that yields a {\it quadratic} equation for $w^3$,
\be \lb{quadrvi}
(w^3)^2 - \fracmm{q^2}{8}w^3 +\fracmm{r^3}{27}=0~~,
\ee
whose roots are
\be \lb{qvieter}
w^3_{1,2} = (q/4)^2 \left[ 1 \pm \sqrt{ 1 - \fracmm{(r/3)^3}{(q/4)^4} } \right] ~~.
\ee

Going back to the critical condition (\ref{hpotcr}) in the form
\be \lb{hpotcr2}
F^2 = e^{6\c_c}\left[  1 - \fracmm{1}{3}a^{-2}e^{2\c_c}\right]~,
\ee
and inserting it into the potential (\ref{hpot}) yields the {\it height} of the inflationary potential 
$V_{\rm plateau}$ at the onset of inflation,
\be \lb{plateau}
M_{\rm Pl}^{-4}V_{\rm plateau}=e^{-6\c_c}\left[  \fracmm{2}{3}a^{-2}e^{2\c_c}-1\right]~.
\ee
Demanding its positivity, $V_{\rm plateau}>0$, gives us the restriction  (\ref{stab}) again.

The second derivative of the potential (\ref{hpot}) at the critical point  (\ref{hpotcr}) is given by
\be \lb{secdc}
\left. \fracmm{\pa^2 V}{\pa \c^2} \right|_{\c=\c_c}=8e^{-6\c_c}\left( 9 -4a^{-2} e^{2\c_c}\right)~.
\ee
Its positivity (stability) implies
\be \lb{posh}
\fracmm{F^2}{\tilde{\g}_8}<54~~.
\ee
Taken together with (\ref{stab}) and (\ref{dicsp}), this implies that the values of the ratio $F^2/\tilde{\g}_8$ have to be restricted as follows:
\be \lb{rangerat}
\eqalign{
16<\fracmm{F^2}{\tilde{\g}_8}\leq 27~~, & \qquad \D_{3,4} \leq 0~~,\cr
27\leq \fracmm{F^2}{\tilde{\g}_8}< 54~~, & \qquad \D_{3,4} \geq 0~~.\cr}
\ee

Because of $1<F^2/(16\tilde{\g}_8)\equiv 1+\d<(\frac{3}{2})^3$, it is instructive to investigate the case of $0<\d\ll 1$ describing the strong stabilization of the modulus $\c$.  In this case,  (\ref{smvac}) and (\ref{hpotcr2})  give rise to
\be \lb{difdelta} 
 0<\c_c-\c_0 \approx \fracmm{1}{12}\d \ll 1~~,
\ee
leading to a {\it single-field inflation} driven by inflaton (scalaron) $f$ indeed.

The physical {\it hierarchy} of scales ({\it cf.} Eq.~(\ref{hier})) reads
\be \lb{hier2}
 m\low{\hat{f}_0} <  m\low{\hat{\c}_0} \ll M\low{\rm KK} \ll  M\low{\rm Pl}~~.
\ee
The KK scale in our case is given by $M_{\rm KK}\approx e^{-\c_0}M_{\rm Pl}$, where the presence
of the warp factor is dictated by the compactification ansatz (\ref{mansatz}).

The condition $M\low{\rm KK} \ll  M\low{\rm Pl}$ implies
\be \lb{firstc}
2F^2 \gg 1
\ee
because of (\ref{smvac}). The condition  $m_{\hat{\c}_0} \ll M\low{\rm KK}$ implies 
\be \lb{secondc}
F^2 \gg \sqrt{2}
\ee
that is slightly stronger than (\ref{firstc}). Both conditions can be easily satisfied by taking
 $F^2\gg 1$.

The remaining condition  $m\low{\hat{f}_0} <  m\low{\hat{\c}_0}$ implies $F^2/\tilde{\g}_8 <72$ that is
already satisfied under the conditions (\ref{rangerat}). However, it is not possible to get 
 $m\low{\hat{f}_0} \ll  m\low{\hat{\c}_0}$ here.

\section*{Appendix D : $D=8$ gauged supergravity}

The $D=8$ supergravity (with 16 supercharges) received relatively little attention in the literature versus the supergravities in $D=10$ and $D=11$. For our purposes, we need a $D=8$ supergravity modified by the quartic scalar curvature term and having a scalar potential. In this Appendix we recall the $SU(2)$ gauged $N=2$ supergravity in $D=8$, which was derived by Salam and Sezgin \cite{ss8} by using the
Scherk-Schwarz-type dimensional reduction \cite{ssgred} of the 11-dimensional supergravity \cite{cjs}.

The 11-dimensional supergravity \cite{cjs} is {\it unique}, so that it is the good point to start with. Its standard action is well known, while its existence can be related to the existence of the 11-dimensional supermultiplet containing the 11-dimensional spacetime scalar curvature $\mathscr{R}$ among its field components. Therefore, there is little doubt that the $(\mathscr{R}+\mathscr{R}^4)$ {\it supergravity} action in $D=11$ also exist, though (to the best of our knowledge) it was never constructed in the literature. So, assuming its existence, we write down the relevant part of its bosonic terms as~\footnote{We use the spacetime signature $(-,+,+,...,+)$ in $D=11$.}
\begin{equation} 
S_{11}=\int d^{11}X\fracmm{E}{2\tilde{\kappa}^2}\left(\mathscr{R}+\tilde{\gamma}\mathscr{R}^4-\fracmm{\tilde{\kappa}^2}{12}G_{ABCD}G^{ABCD}+\fracmm{8\tilde{\kappa}^3}{144^2}\varepsilon^{A_1...A_{11}}G_{A_1...}G_{A_5...}V_{...A_{11}}\right)~,\label{S11}
\end{equation}
where we have simply added the quartic curvature term (with the coupling constant $\tilde{\gamma}$) to the standard bosonic action of the 11-dimensional supergravity. Of course, adding the $\mathscr{R}^4$ term also requires adding its supersymmetric completion that is going to result in more bosonic terms in the action. However, because all extra terms are going to be the higher-derivative couplings of the bosonic 3-form field, also non-minimally coupled to gravity, we assume that these extra couplings are irrelevant for the {\it scalar} sector of the theory (see below).~\footnote{It is worth mentioning that our approach is apparently {\it different} from M-theory, because we treat the $\mathscr{R}^4$ term nonperturbatively, so
that its presence leads to the new physical degrees of freedom in $D=11$, which are absent in the
standard $D=11$ supergravity, similarly to the $(R+R^4)$ gravity in lower dimensions.}

As regards our notation, we denote $E\equiv\det {E_M}^A$ in terms of an {\it elfbein} 
${E_M}^A$ in $D=11$. Here we denote  the 11-dimensional Lorentz indices by early capital latin letters as $A,B,C,...$, while the middle capital latin letters $M,N,P,...$ are used for  the 11-dimensional Einstein (curved) indices. The $\tilde{\kappa}$ is the gravitational constant in $D=11$. The scalar curvature $\mathscr{R}$ is defined in terms of the spin connection 
\begin{gather}
\omega_{ABC}\equiv E^M_A\omega_{MBC}=\frac{1}{2}\eta_{CE}(E^M_AE^N_B-E^M_BE^N_A)\partial_ME_N^E-\frac{1}{2}\eta_{AE}(E^M_BE^N_C-E^M_CE^N_B)\partial_ME_N^E\nonumber\\+\frac{1}{2}\eta_{BE}(E^M_CE^N_A-E^M_AE^N_C)\partial_ME_N^E \label{spinc}
\end{gather}
as
\be \lb{sc11}
\mathscr{R}=\omega_{ABC}\omega^{CAB}+\omega_A\omega^A-2E^{-1}\partial_M(E{E^M}_A\omega^A)~,
\end{equation}
where $\omega_A\equiv \eta^{BC}\omega_{BCA}$ and $\eta_{AB}$ is Minkowski metric in $D=11$. The 
4-form field strength $G_{ABCD}$ is defined in terms of the 3-form gauge potential $V_{ABC}$ as
\begin{equation} \lb{gp11}
G_{ABCD}=4\partial_{[A}V_{BCD]}+12{\omega_{[AB}}^E V_{CD]E}~.
\end{equation}


To dimensionally reduce the modified $D=11$ supergravity to eight dimensions on a sphere $S^3$, we use the ansatz \cite{ss8}
\begin{equation}
{E_M}^A=
 \begin{pmatrix}
  e^{-\tilde{\kappa}\phi/3}e^a_\mu & 0 \\
  2\tilde{\kappa} e^{2\tilde{\kappa}\phi/3}A^\alpha_\mu L^i_\alpha & e^{2\tilde{\kappa}\phi/3}L^i_\alpha
 \end{pmatrix}~,\quad
{E^M}_A=
 \begin{pmatrix}
  e^{\tilde{\kappa}\phi/3}e_a^\mu & -2\tilde{\kappa} e^{\tilde{\kappa}\phi/3}e^\mu_a A^\alpha_\mu\\
  0 & e^{-2\tilde{\kappa}\phi/3}L_i^\alpha
 \end{pmatrix}~,\label{ansatz8}
\end{equation}
where we have introduced the 8-dimensional Lorentz indices $a,b,c,...$ and the 8-dimensional Einstein indices
 $\mu,\nu,\rho,...$, as well as the 3-dimensional (compact) Lorentz and Einstein indices, $i,j,k,...$ and $\alpha,\beta,\gamma,...$, respectively. The dilaton $\phi$ represents the volume modulus of the 3-sphere, the  $e_\mu^a$ is an 8-dimensional {\it achtbein}, the $L^i_\alpha$ is the unimodular matrix ($\det L^i_\alpha=1$) having 5 scalars, and the $A_\mu^\alpha$ is a set of 8-dimensional vectors. 

The Scherk-Schwarz dimensional reduction is used to gauge symmetries of a compact manifold in the reduced theory by allowing the fields to depend on the compact coordinates
\cite{ssgred}. Let us denote the non-compact coordinates by $\{x\}$, and the compact  coordinates by
 $\{y\}$, and then factorize the $y$-dependence as
\be \lb{factxy}
e^a_\mu(x,y)=e^a_\mu(x)~,\quad
A^\alpha_\mu(x,y)={U^{-1\alpha}}_\beta(y)A^\beta_\mu(x)~,\quad
L^i_\alpha(x,y)={U_\alpha}^\beta(y)L^i_\beta(x)~,
\ee
where ${U_\alpha}^\beta(y)$ are elements of the gauge group $SU(2)$ in our case. The $SU(2)$ structure constants 
\begin{equation}
f^\gamma_{\alpha\beta}\equiv {U^{-1}_\alpha}^{\alpha'}{U^{-1}_\beta}^{\beta'}(\partial_{\beta'}{U_{\alpha'}}^\gamma-\partial_{\alpha'}{U_{\beta'}}^\gamma)=-\fracmm{g}{2\tilde{\kappa}}\varepsilon_{\alpha\beta\delta}g^{\delta\beta}
\end{equation}
are $y$-{\it independent}, where we have introduced the $SU(2)$ gauge coupling constant $g$ and 
the 3-dimensional Levi-Civita tensor $\varepsilon_{\alpha\beta\gamma}$.

Substituting the ansatz \eqref{ansatz8} into \eqref{spinc} reduces the spin connection components as
\cite{ss8}
\begin{align}
\begin{aligned}
&\omega_{abc}=e^{\tilde{\kappa}\phi/3}(\tilde{\omega}_{abc}-\frac{1}{3}\tilde{\kappa}\eta_{ab}\partial_c\phi+\frac{1}{3}\tilde{\kappa}\eta_{ac}\partial_b\phi)~,\\
&\omega_{abi}=\tilde{\kappa} e^{4\tilde{\kappa}\phi/3}F_{abi}~,\\
&\omega_{aij}=e^{\tilde{\kappa}\phi/3}Q_{aij}~,\\
&\omega_{iab}=-\tilde{\kappa}e^{4\tilde{\kappa}\phi/3}F_{abi}~,\\
&\omega_{ija}=e^{\tilde{\kappa}\phi/3}(P_{aij}+\frac{2}{3}\tilde{\kappa}\delta_{ij}\partial_a\phi)~,\\
&\omega_{ijk}=-\frac{g}{4\tilde{\kappa}}e^{-2\tilde{\kappa}\phi/3}(\varepsilon_{jkl}{T_i}^l+\varepsilon_{kli}{T_j}^l-\varepsilon_{lij}{T_k}^l)~,\label{spincc}
\end{aligned}
\end{align}
where we have used the notation 
\begin{gather}
F^\alpha_{\mu\nu}=\partial_\mu A_\nu^\alpha-\partial_\nu A_\mu^\alpha+g\varepsilon_{\alpha\beta\gamma}A^\beta_\mu A^\gamma_\nu~,\nonumber \\
P_{\mu ij}+Q_{\mu ij}=L_i^\alpha(\delta^\beta_\alpha\partial_\mu-g{{\varepsilon_\alpha}^\beta}_\gamma A^\gamma_\mu)L_{\beta j}~,\label{PQ} \\
T^{ij}=L^i_\alpha L^j_\beta \delta^{\alpha\beta}~,\nonumber
\end{gather}
with $P_{\mu ij}$ representing the symmetric part of the r.h.s. of \eqref{PQ}, and $Q_{\mu ij}$ representing the antisymmetric part. The fields $L_\alpha^i$ are subject to the relations \cite{ss8}
\begin{equation}
L^i_\alpha L^j_\beta\delta_{ij}=g_{\alpha\beta},\;\;L^i_\alpha L^j_\beta g^{\alpha\beta}=\delta^{ij}~,
\end{equation}
where $g_{\alpha\beta}$ is the metric of the compact manifold $(S^3)$.

As regards $V_{ABC}$ and $G_{ABCD}$, their relevant components are
\begin{gather}
\begin{gathered}
\varepsilon_{\alpha\beta\gamma}B\equiv e^{2\tilde{\kappa}\phi}L^i_\alpha L^j_\beta L^k_\gamma V_{ijk}~,\\
\varepsilon_{\alpha\beta\gamma}\partial_\mu B\equiv e^{5\tilde{\kappa}\phi/3}e^a_\mu L^i_\alpha L^j_\beta L^k_\gamma G_{aijk}~,\label{GV}
\end{gathered}
\end{gather}
where $B$ is another {\it scalar} field.

Equations \eqref{ansatz8}, \eqref{spincc}, and \eqref{GV} allow us to rewrite the 11-dimensional action \eqref{S11} as
\be \lb{11act2}
\eqalign{
S_{11}= & \int d^8x d^3y\;U(y)\fracmm{e}{2\tilde{\kappa}^2}[R+\tilde{\gamma}e^{2\tilde{\kappa}\phi}R^4-\tilde{\kappa}^2e^{2\tilde{\kappa}\phi}F_{\mu\nu}^\alpha F^{\mu\nu}_\alpha-2\tilde{\kappa}^2\partial_\mu\phi\partial^\mu\phi \cr
&  -\fracmm{g^2}{4\tilde{\kappa}^2}e^{-2\tilde{\kappa}\phi}(T_{ij}T^{ij}-\frac{1}{2}T^2)\\
-P_{\mu ij}P^{\mu ij}-\frac{1}{2}\tilde{\kappa}^2e^{-4\tilde{\kappa}\phi}\partial_\mu B\partial^\mu B \cr
& -\fracmm{\tilde{\kappa}^2}{12}e^{2\tilde{\kappa}\phi}G_{\mu\nu\rho\sigma}G^{\mu\nu\rho\sigma}-\fracmm{\tilde{\kappa}^3}{432}e^{-1}\varepsilon^{\mu_1...\mu_8}G_{\mu_1...\mu_4}G_{\mu_5...\mu_8}B]+\ldots~, \cr}
\ee
where $U(y)\equiv \det {U_\alpha}^\beta(y)$, $T\equiv {T_i}^i$, $R$ is the $8$-dimensional scalar curvature and  the dots stand for irrelevant terms. Since the only $y$-dependent function is $U(y)$, one can perform $y$-integration with
\begin{equation}
\int d^3y U(y)=V_0~,
\end{equation}
defining the invariant volume $V_0$ of the compact manifold $(S^3)$. With the gravitational coupling 
$\kappa=\tilde{\kappa}/\sqrt{V_0}$ in $D=8$, rescaling dilaton as $\phi\rightarrow\phi/\sqrt{V_0}$ (and similarly for the other fields $A_\mu^\alpha$, $B$ and $V_{\mu\nu\rho}$) and rescaling the gauge coupling 
as $g\rightarrow g\sqrt{V_0}$ leads to the action 
\begin{gather} 
S_8=\int d^8x\fracmm{e}{2\kappa^2}[R+\tilde{\gamma}e^{2\kappa\phi}R^4-\kappa^2e^{2\kappa\phi}F_{\mu\nu}^\alpha F^{\mu\nu}_\alpha-2\kappa^2\partial_\mu\phi\partial^\mu\phi-V(T)-P_{\mu ij}P^{\mu ij}\nonumber\\
-\fracmm{1}{2}\kappa^2e^{-4\kappa\phi}\partial_\mu B\partial^\mu B-\fracmm{\tilde{\kappa}^2}{12}e^{2\kappa\phi}G_{\mu\nu\rho\sigma}G^{\mu\nu\rho\sigma}-\fracmm{\kappa^3}{432}e^{-1}\varepsilon^{\mu_1...\mu_8}G_{\mu_1...\mu_4}G_{\mu_5...\mu_8}B]+...~,\label{8saction}
\end{gather}
whose scalar poitential is given by \cite{ss8}
\begin{equation} \lb{spot8}
V(T)=\fracmm{g^2}{4\kappa^2}e^{-2\kappa\phi}(T_{ij}T^{ij}-\frac{1}{2}T^2)~.
\end{equation}

\vglue.2in

\end{document}